\documentclass[a4paper,fleqn,usenatbib]{mnras}

\usepackage{newtxtext,newtxmath}

\usepackage[T1]{fontenc}
\usepackage{ae,aecompl}

\usepackage{verbatim}

\usepackage{graphicx}	
\usepackage{amsmath}	
\usepackage{amssymb}	
\usepackage{multicol}        

\title[A tale of two clump masses]{A Tale of Two Clump Masses: A new way to study clump formation in simulations}

\author[S. M. Benincasa et al.]{
S. M. Benincasa,$^{1,2}$\thanks{E-mail: smbenincasa@ucdavis.edu}
J. W. Wadsley,$^{1}$
H. M. P. Couchman,$^{1}$
A. R. Pettitt$^{3}$
\newauthor
and E. J. Tasker$^{4}$
\\
$^{1}$Department of Physics \& Astronomy, McMaster University, 1280 Main St. W, Hamilton L8S 4K1, Canada \\
$^{2}$Department of Physics, University of California, Davis, CA 95616, USA \\
$^{3}$Department of Physics, Faculty of Science, Hokkaido University, Sapporo 060-0810, Japan\\
$^{4}$Institute of Space and Astronautical Science, Japan Aerospace Exploration Agency, Yoshinodai 3-1-1, Sagamihara, Kanagawa 252-5210, Japan
}

\date{Accepted XXX. Received YYY; in original form ZZZ}

\pubyear{2015}

\begin{document}
\label{firstpage}
\pagerange{\pageref{firstpage}--\pageref{lastpage}}
\maketitle

\begin{abstract}
We present a new method to study the characteristic scales of collapse and fragmentation in galactic disks.  Clump formation is seeded in simulations via controlled perturbations with a specified wavelength and velocity.  These are applied to otherwise quiet gas disks ranging from analogues of present day spirals to gas-rich, high-redshift galaxies.  The results are compared to linear theory, turbulently perturbed disks and observations.  The results reflect the expectations of linear, non-axisymmetric theory with a finite window for growth into a bound clump.  We identify two new modes of clump formation: rotation-driven fission and fragmentation of tidal tails, though both are expected to rarely contribute to clump formation in observed disks.  We find that bound clumps are generally much smaller than the commonly used Toomre mass.  The preferred scale for fragmentation increases with the disk gas mass but cannot produce bound objects larger than $\sim10^9$ M$_{\odot}$.  The most likely bound clump mass increases from $3\times10^6$ in low mass disks up to $5\times10^8$ M$_{\odot}$.  We conclude that observed massive stellar and gaseous clumps on 1 kpc scales at high redshift are most likely aggregates of many initially distinct bound clumps.
\end{abstract}



\section{Introduction}
The typical size of star clusters is expected to depend on the galactic environment. The Jeans' mass, which changes based on environment, plays a role \citep[e.g.][]{hopkinse2012}. The pressure likely plays a role; high pressure environments such as Arp 220 have larger star clusters when compared to other local galaxies \citep[e.g.][]{wilson2006}. If we begin by considering low redshift galaxies, a preferred mass-scale for star cluster formation is apparent.  In the Milky Way itself, star clusters have typical masses between $10^3 - 10^4$ M$_{\odot}$ \citep{fall2012}. A preferred scale for star clusters may in turn suggest a preferred scale for Giant Molecular Clouds (GMCs).  In the Milky Way GMCs have typical masses between $10^5 - 10^6$ M$_{\odot}$ and typical sizes of 50 - 100 pc \citep{fukuiARAA, mckeeARAA}.  
 
If we consider galaxies at higher redshifts, these preferred scales appear to change. Stellar observations are able to identify large UV-bright star-forming regions called {\it clumps} \citep{elmegreen2007}.  The CANDELS survey has provided extensive clump catalogues for galaxies between 0.5 $< z <$ 3.5.  The properties of clumps identified in CANDELS galaxies suggest they are extremely large, with typical masses of $10^7 - 10^9$ M$_{\odot}$ and typical sizes $\sim 1$ kpc \citep{CANDELS, CANDELS2}.  Other compilation studies find masses between $10^5 - 10^9$ M$_{\odot}$ \citep{dessauges2018}.  Either way, these are orders of magnitude more massive than present-day clusters or star-forming regions.

This picture is even more complex if we add in starbursts or merger-driven systems.  As mentioned above, if we consider Arp 220, there are many active sites of star formation and the star-forming complexes may be much larger than those in the Milky Way \citep[][and references therein]{murray2010}.  In these environments the masses of stellar clusters increase dramatically.  \citet{wilson2006} find masses approaching $10^7$ M$_{\odot}$, which may make them candidates for young globular clusters.

The conditions in star-forming regions should be imprinted in the properties of gas. At higher redshifts, and in starburst systems, the conditions both within and around galaxies were different than at low redshift. Specifically, at higher redshifts the galaxy interaction rate was higher and galaxies themselves are much more likely to be gas rich.

It is no surprise that in such gas-rich, highly molecular environments the properties of star-forming regions are likely to be different.  Indeed, the appearance of gas disks beyond z$\sim$0.5 are much more clumpy in nature \citep[e.g.][]{forster2009}.  This highly molecular clumpy nature may suggest that star-forming regions may be larger in both mass and spatial extent. Wide beam observational studies suggest these objects have masses between $10^8$ - $10^{10}$ M$_{\odot}$, or approximately $1-10\%$ of the total disk mass \citep[e.g.][]{tacconi2010, swinbank2010, swinbank2011, genzel2011, hodge2012}.  The corresponding physical sizes range from as small as 100 pc to as large as 2 kpc \citep[e.g.][]{swinbank2010, tacconi2010}. 

However, recent results from the SGASS lensing survey have shown a finer level of substructure, albeit in galaxies less massive than those typical in the CANDELS sample \citep{johnson_model}.  These results show that stellar clump sizes can be consistent with present-day star clusters, which would originate from objects similar to present day GMCs with high star formation efficiency \citep{johnson2017, rigby2017}.   Another such lensing study has been done in the Cosmic Snake \citep{cava2018}.  Using data from the CLASH survey, the authors have obtained both a lensed arc and counterimage.  In this way they can compare two spatial resolutions for the same object.  They find that the lower resolution image (counterimage) produces clumps that are amplified by a factor of 2-5 on average, for an decrease in resolution of 10 times.  Studies like these suggest two important points. First, the gas mass of a galaxy is important for determining the scale of star formation.  Second, the resolution of earlier studies may not be sufficient to resolve clumps.

With the typical physical resolution of instruments at high redshift being generally poorer, it may be that we are treating collections of GMCs \citep{tacconi2010} as a single entity.  This idea has been lent credence by samples of lensed galaxies: while directly observed galaxies are often large due to selection effects, lensed galaxies are typically of lower mass (M$_{\star}$ $\sim 10^9$ M$_{\odot}$).  They provide us with a better resolved picture of the molecular gas.  For example, \cite{swinbank2010} find gas clumps of similar size to Milky Way GMCs, approximately 100 pc.  They propose that these objects would be similar to present-day star-forming regions except with more star-forming cores with higher densities.  \cite{hodge2012} infer typical internal densities of $\sim100$ cm$^{-3}$, in accordance with the typical density of low-redshift GMCs.
If we infer masses from these sizes, we can assume that these objects would likely have masses similar to large present-day GMCs: maybe between $10^5-10^7$ M$_{\odot}$.
Local starburst galaxies with enhanced star-formation and kinematic properties similar to galaxies at z $\sim$ 1.5 can also be used as a testbed for these theories.  Their proximity offers better resolution studies and HST-DYNAMO has studied 13 such galaxies \citep{white2017}.  Studies here confirm that clump clustering is likely to impact the measurement of clump properties at higher redshifts, where resolution degrades \citep{fisher2017b}.  

If we look to isolated galaxy simulations, we see results consistent with small star-forming regions.  For example, \citep{tamburello2015} find smaller clump masses ($< 10^7$ M$_{\odot}$).  The higher resolution available in isolated galaxy simulations also offers the perfect place to study the impacts of resolution on structure identification.  Some work has suggested that the resolution of HST at the redshifts concerned is insufficient to fully resolve these clumpy objects.  \cite{tamburello2017} and \cite{dessauges2017} have argued that at the  spatial resolution of 1 kpc, it is not possible to fully resolve these stellar clumps.  This supports the idea that we are just seeing collections of smaller stellar clumps, clustered closely together.  Indeed, \cite{behrendt2016} have shown that closely clustered clumps could be confused for more massive objects with the resolution of surveys like CANDELS.

At the other extreme, it has been suggested that in massive, gas-rich galaxies the physics of clump formation changes with Violent Disk instabilities (VDI) producing different outcomes \citep{dekel2009}.  This is built on the idea that a Toomre instability sets the characteristic scale of clumps in gas-rich disks \citep{shlosman1993,noguchi1998,noguchi1999}.  This idea has been invoked to explain certain cosmological zoom simulations exhibiting larger clumps in the range 10$^{7-9}$ M$_{\odot}$ \citep{mandelker2014, mandelker2017}.  However, it has been theorized that not all gas rich disks host VDI, and that this behaviour is largely dependant on feedback strength rather than a qualitative difference in how clumps form \citep{fiacconi2017}.  Regardless, other studies of cosmological zoom simulations also report clump masses in this higher range \citep{agertz2009,oklopcic2017}.  

An alternative way to study clump formation in isolated galaxies is by employing spherical halo collapse. Here, rather than begin with a disks, a halo of dark matter and hot gas are allowed to collapse, thus forming the disk \citep[e.g.][]{kaufmann2006, kaufmann2007, teyssier2013}. This is commonly used to study the interaction of the disk with the surrounding medium, to study cold flows for example. However, it has also been used to study clump formation \citep{noguchi1999, inoue2012}. For example, \citet{inoue2014} report clump masses above $10^8$ M$_{\odot}$ in isolated galaxies formed in this way.


Resolving these differences is complicated by the fact that all of these studies use different methods.  Some involve simulations on cosmological scales while others model isolated galaxies, the resolution of these two types of simulations can be quite different.  Different studies use different numerical methods or different hydrodynamical schemes.  Beyond that, perhaps the largest variable, is the type of feedback chosen. The type and strength of feedback chosen plays a large role in determining the structure of star-forming gas and, consequently, the structure of stellar clusters.  All of these variables makes it incredibly confusing to compare results among different studies.  Add on top of that the different types of observations we are considering, stellar versus gas, lensed versus un-lensed, and we are left with great difficulty in interpreting the results in the literature.

We propose a new method to study clump formation in simulations. Our method avoids the problems associated with many of the algorithm-specific assumptions discussed above.  We seed clump formation events by hand and study their growth in high resolution isothermal disks that do not include feedback.  In this way, we can constrain the initial mass of clumps formed in a variety of disks.  These are directly comparable to both observationally determined masses and masses from theories of fragmentation.

The rest of the paper is laid out as follows.  We begin by examining the predictions from linear theory in section 
~\ref{sec:theory}.  Linear theory is difficult to extrapolate to non-linear clumps properties.  Instead, we use it to design simulations to explore clump formation in disks ranging from Milky Way-like cases to the heavy, turbulent disks expected at high redshifts.   In section~\ref{sec:methods} we describe our controlled simulation approach which allows for high resolution and relatively easy interpretation of the results.   We make first use of our quiet disks in section ~\ref{sec:turbdisk}, looking at isolated, turbulent disks. These simulations show typical outcomes for turbulent-type initial conditions. Their purpose in this paper is to illustrate the value of a more controlled approach.  Finally, in order to study the key scales for fragmentation and clump formation in a controlled way, we take a new approach of seeding non-linear perturbations.  We present details of the approach and results in section~\ref{sec:seed}.  In section~\ref{sec:mass}, we extrapolate from our simulation results to estimate likely clump masses based on disk conditions.  Finally, in section~\ref{sec:implications} we discuss the observational implications of this study. 

\section{Theoretical Expectations}\label{sec:theory}

While galactic disks are complex systems they are still amenable to theoretical analysis.  The classic analysis by \cite{toomre1964} assumed a razor-thin, axi-symmetric system.   This analysis applies in the case where the perturbations are effectively rings or very tightly wound (highly localized).   In this case the dispersion relation has three main terms,
\begin{equation}
   \omega^2 = \kappa^2 - 2 \pi G \Sigma |k| + {c_s}^2 k^2.
\end{equation}
The terms representing rotation (the epicycle frequency, $\kappa$) and pressure (the sound speed, $c_s$) act to stabilize the perturbations against gravity (Newtonian constant $G$) due to the underlying surface density $\Sigma$.  As the physical scale, given by the wavenumber $k$, changes from large ($k\sim$0) to small, we transition from being stabilized by rotation to sound waves or pressure; a stabilized regime is one in which the oscillation frequency, $\omega^2 > 0$. A key physical scale is the Toomre length,
\begin{equation}
   \lambda_{\rm Toomre} = \frac{4 \pi^2 G \Sigma}{\kappa^2}.\label{eqn:disp}
\end{equation}
Beyond this scale all perturbations are stabilized by rotation.  This sets a hard upper limit on the mass of clumps collapsing directly from a single perturbation.  This can be translated into a mass, the Toomre mass, by assuming intrinsically circular collapsing regions,
\begin{equation}
M_{\rm Toomre} = \pi \left(\frac{\lambda_{\text{Toomre}}}{2}\right)^2 \Sigma. 
\end{equation}
This mass is sometimes used an initial mass for clumps \citep{reina2017, kruijssen2012}.
However, this is somewhat ad hoc given that the linear dispersion relation applies to plane waves which would collapse to filaments or rings in the case of axi-symmetric global perturbations.  So based on linear, axi-symmetric theory it is, at best, a rough guide.  

At intermediate scales, $\lambda \sim \lambda_{\rm Toomre}/2$, gravity is at its most effective relative to stabilizing forces and $\omega$ can be imaginary (unstable) if the Toomre $Q$ parameter is less than one,
\begin{equation}
   Q = \frac{c_s \kappa}{\pi G \Sigma} < 1.
\end{equation}
This critical scale of $\lambda_{\rm Toomre}/2$ is a starting point for a characteristic scale for fragmentation.
The aforementioned violent disk instabilities are effectively non-linear Toomre instabilties.   Thus the Toomre $Q$ parameter should be a guide to locations where gas and stellar clumps can form \citep{inoue2016}.   It has been suggested these instabilities are directly involved in setting clump masses until z$\sim$1-0.5 \citep{cacciato2012}.

There are several complicating factors with applying these results directly to galactic disks.  The first is that real disks are not razor thin.  The primary impact is that disks can have $Q$ as low as $\sim 2/3$ without being unstable to axi-symmetric modes depending on how thick the disk is \citep{romeo2011}.  Secondly, galaxies are comprised of both gas and stars which act together and affect each other's individual stability \citep{goldreich1965, romeo2010, agertz2015}.    Another limiting complication is that the linear theory is a local approximation and does not apply for structures that are comparable in size to the disk.

A key feature of these linear modes is that the growth rate is independent of both the amplitude and time so that for $\omega^2 < 0$ they are predicted to grow indefinitely so that finite disturbances may result.   In other words, perturbations can grow indefinitely without being limited by rotation.

A less commonly considered factor is the assumption of axi-symmetry or tight winding.   It is widely recognized that disks grow large-scale perturbations (e.g. spiral structure) for supposedly stable $Q$-values in the range of 1-2.  Secondly, the structures that grow are not axisymmetric.   Axi-symmetry greatly simplifies the dynamics as such perturbations do not elongate due to shear.   Local non-axi-symmetric perturbations including the role of rotation and shear were examined by \citet{goldreich1965b}.  In this case, the behaviour of plane waves does not simplify to a quadratic dispersion relation and even linear waves must be integrated as differential equations.  Here we present the evolution equation for small amplitude surface density perturbations in the form given by \citet{jog1992}, and simplified for the case of a single gaseous component,
\begin{multline}
  \left(\frac{d^2 \theta}{d\tau^2}\right) - \left(\frac{d\theta}{d\tau}\right)\left(\frac{2\tau}{1+\tau^2}\right) = \\ \frac{-\theta}{4 A^2}\left[\kappa^2 + \frac{8 A B}{1+\tau^2} - 2 \pi G \Sigma k_y \sqrt{1+\tau^2} + c_s^2 k_y^2(1+\tau^2) \right],
\label{eqn:glb}
\end{multline}
where $\theta = \delta\Sigma/\Sigma$ is the fractional perturbation to the gas surface density, $\tau = 2 A t - k_x/k_y$ is a dimensionless time (such that the perturbation is radial at $\tau=0$), $\Sigma$ is the unperturbed surface density, $\kappa$ is the epicycle frequency, $A$ and $B$ are the Oort constants, $(k_x, k_y)$ is the wavevector, and $c_s$ is the gas sound speed.

The terms in the square brackets are directly analogous to the terms in the axi-symmetric dispersion relation, Eqn.~\ref{eqn:disp},  and reduce to it in the appropriate limit.   The first two terms represent shear and rotation but with an added time dependence associated with the instantaneous orientation of the wavefront.  The latter terms depend on the instantaneous wavenumber which is sheared and thus minimized near $\tau=0$.  The pressure term always dominates for very early and late times.

Thus the behaviour of these linearized equations (in $\theta$) is such that growth occurs briefly near $\tau \sim 0$ as the waves transition from leading to trailing, as noted by \citet{goldreich1965b} and \citet{jog1992}.   The behaviour is oscillatory for other times.  This means that the net growth is limited.   Thus to achieve interesting outcomes from a relatively quiet start we must have multiple cycles of growth enabled by non-linear (e.g. wave to wave) interactions that reform leading waves.  Put another way, to achieve bound structures we must start with substantial perturbations and grow these.

In the vigorously star forming disks of interest, we have feedback and turbulence to provide non-linear perturbations.  For example, individual superbubbles sweep up material on kpc scales.  Interactions of many feedback events are expected to result in a turbulent velocity spectrum over a range of scales up to of order the disk scale height.  Structure in the disk, such as pre-existing spiral modes, can also extract energy from the disk rotation to power turbulence on scales similar to the spiral waves themselves; this is comparable to the disk size for very unstable disks.  Thus we would argue from the theory that the growth of finite, non-linear perturbations is a consistent picture of the development of bound clumps.

\begin{table*}
	\centering
	\caption{The three main disk initial conditions discussed in this work.}
	\label{tab:example_table}
	\begin{tabular}{lccccccc} 
		\hline
		name & $c_s$ & T & mass & $\Sigma_g$(R $= 5$ kpc) & $Q_{\text{min}}$ & $\lambda_{\text{Toomre}}$ & $M_{\rm Toomre}$ \\
		\hline
		cold disk & 5.16 km/s & 2 500 K & 6.95 $\times 10^9$ M$_{\odot}$& 22 M$_{\odot}$/pc$^2$ & 1.07 & 960 pc & 1.5 $\times 10^7$ M$_{\odot}$\\
		warm disk & 14.59 km/s & 20 000 K & 1.96 $\times 10^{10}$ M$_{\odot}$ & 61.6 M$_{\odot}$/pc$^2$ & 1.15 & 2.4 kpc & 2.9  $\times 10^8$ M$_{\odot}$\\
		hot disk & 35.75 km/s & 120 000 K & 5.56 $\times 10^{10}$ M$_{\odot}$ & 176 M$_{\odot}$/pc$^2$ & 1.1 & 5.4 kpc & 4.1  $\times 10^9$ M$_{\odot}$\\
		\hline
	\end{tabular}
	\label{tab:summary}
\end{table*}

\section{Simulation Methods and Disk Models}\label{sec:methods}

Astrophysical simulations always struggle to resolve the turbulent cascade and small scale structures developed through the coupling of turbulence and thermal instabilities.   Taken from a broad perspective, however, turbulence behaves similarly to a polytropic gas.  For this work, we have used isolated disks with an isothermal equation of state which is very simple to model relative to the full complexities of small scale turbulence.  However, it directly provides the effective support we require of the turbulence and simultaneously provides a simple, straightforward model for the cooling losses.   The effective equation of state of the ISM is complex but broadly similar to an isothermal one.  In particular, \cite{goldreich1965b} show that for equations of state with polytropic index $\gamma \sim 1$, non-linear unstable clumps will remain unstable and continue to collapse.  Thus with this choice we can have confidence regarding the future fate of collapsing regions. 

All of the disk simulations presented here use a static, logarithmic halo potential \citep{BT} to represent the background of the dark matter halo, bulge and old stellar disk.  This provides a well defined rotation curve (i.e. $\kappa(r)$).  In all of the simulations discussed here we set the rotation velocity at $220$ km/s. However, there is a small adjustment to this rotation curve due to the gas component.  The gas surface density is exponential with a scale length of 5 kpc from 2 to 12 kpc with a smooth tailing off to zero below 2 kpc and beyond 12 kpc.  Combined with a constant sound speed, this generates a Toomre $Q$ that varies slowly in the active region from 2 to 12 kpc with a minimum value at $\sim 5$ kpc, and with values exceeding twice the minimum below 2 kpc and beyond 12 kpc.

We use the modern smoothed particle hydrodynamics code \textsc{Gasoline} \citep{wadsley2004, wadsley2017} to simulate the gas component, including self-gravity with a softening length of 10 pc.  We have experimented with applying a Jeans floor that increases the effective pressure where the Jeans length is unresolved \cite[following][]{robertson2008}. However, we find that for the simulations discussed here this makes no difference to the amount of fragmentation.  We do not directly apply a star formation or feedback model as these contain subgrid prescriptions that are code dependent. Additionally, there is still debate regarding the importance of different types of feedback \citep{murray2010, hopkins2012}.

We have built three disk models for this work and their properties are summarized in Table \ref{tab:summary}. These models cover a range of galaxy masses quoted in the literature: from a Milky Way sized galaxy to a galaxy that is massive enough to evolve into a present-day elliptical.

The first is modelled after a Milky Way-type disk, analogous to a less massive high redshift object (which we have labelled cold).  It has  a total gas mass of 6.95$\times 10^9$ M$_{\odot}$ and we use 5.56 million particles.  The gas sound speed is 5 km/s, with a gas particle mass of 1250 M$_{\odot}$.  These smaller mass objects are more typical of lensed samples, as here we are just sampling the galaxy luminosity function.

For the second disk (warm) we have modelled a more massive disk, which is a closer comparison to a high redshift turbulent, gas-rich disk.  This case has a total gas mass of 1.96$\times 10^{10}$ M$_{\odot}$.  Here the gas sound speed is increased to 14 km/s to mimic the larger amount of turbulence present in high redshift galaxies \citep{forster2009}.  To leave the effective resolution and Q parameter the same we then increase the gas particle mass to 3535 M$_{\odot}$.  

Finally, the third disk (hot) is the most massive and warmest of the three.  This case has a total gas mass of 5.56$\times 10^{10}$ M$_{\odot}$.  The gas sound speed is raised to 40 km/s.  Again, to leave the effective resolution the same this results in a gas particle mass of $10^4$ M$_{\odot}$.  Galaxies in this higher mass range are more comparable to samples like that of \cite{CANDELS2} for example.  These galaxies which are already this massive at high redshift will likely become massive ellipticals by the present day.

Since we are interested in triggering clump formation, the surface density profiles chosen are such that the initial Toomre Q parameter lies near the border line of stability.  We choose 5 kpc as the galactic radius of interest, where $Q$ is at a minimum.  At 5 kpc, the cold disk has an initial $Q_{\text{min}}$ of 1.07, the warm disk has an initial $Q_{\text{min}}$ of 1.15, and the hot disk has an initial $Q_{\text{min}}$ of 1.1.  The target value of $Q$ was 1.1 but the effects of self gravity of the gas disk made it hard to get a precise Q, particularly for heavier disks.

For reference, a razor-thin disk stable to axisymmetric, linear perturbations has Q above 1.  In other tests, not reported here, we find that for values above $Q\sim 1.3$, it is difficult for even fairly large perturbations to push regions toward gravitational instability.  Since high-z galaxies have substantial inflows, we expect that they will always evolve to a point with Q between 1 and 1.3 so that clumps can form. It should be noted that none of our disks form clumps or stray from axisymmtery when they are evolved unperturbed.

The Truelove criterion states that in a mesh code four cells are needed to resolve the Jean's length accurately \citep{truelove1997}.  A similar criterion exists in SPH simulations, where we require that the Jean's mass be greater than the neighbour number multiplied by the particle mass \citep{bate1997}. We are able to resolve the Jean's mass up to 100 cm$^{-3}$ in all of our disks, and up to 10$^5$ cm$^{-3}$ two of our three disks. The disk in question is the coldest disk.  We have re-simulated higher resolution cases for a subset of our disks. In these cases we increase the resolution by splitting each gas particle eight times. In these high resolution cases we find no changes to the fragmentation or clump mass.

\begin{figure*}
	\begin{centering}
		\includegraphics[width=\textwidth]{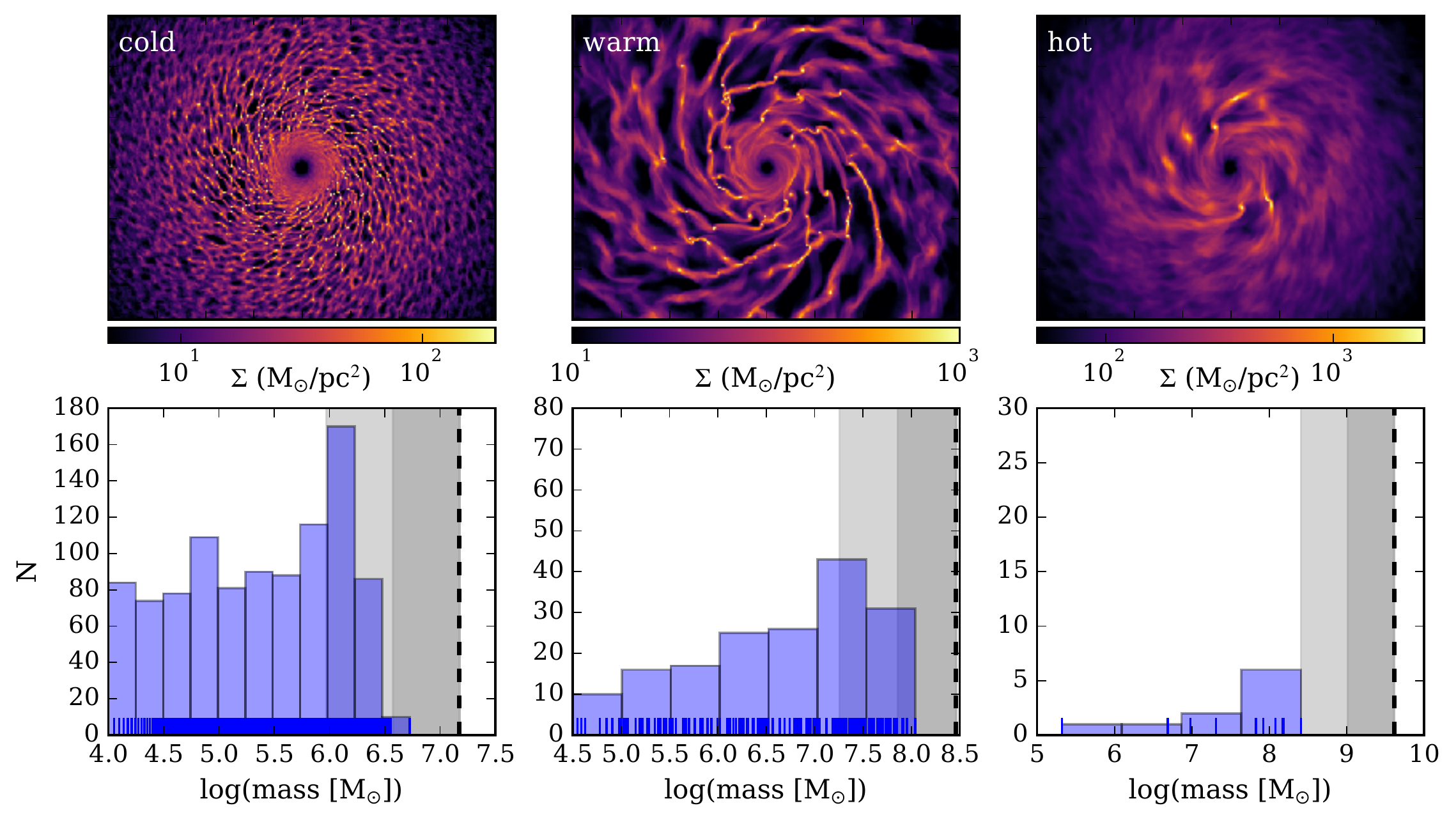}
	\end{centering}
	\caption{Mass distributions for clumps formed in the turbulent disks.  \textit{Top}:  surface density maps of the disks at 40 Myr (this is the time at which all mass measurements are made) These show the entire disk and are 20 kpc across.  \textit{Bottom}: clump mass distributions measured using SKID.  The dashed line shows the Toomre mass for each case.  The light and dark grey shaded regions show the characteristic mass estimated using a radius of $\lambda_{\text{Toomre}/4}$ and $\lambda_{\text{Toomre}/2}$, respectively.  In the distributions, the blue short lines show the masses of each of the clumps to illustrate how the clumps are distributed between the bins.  As the mass of the disk is increased, the mass of the clumps formed likewise increases.  In all cases the clumps formed are less massive than the Toomre mass, sometimes by an order of magnitude.}
	\label{fig:turb}
\end{figure*}

\section{Turbulent Disk Simulations}\label{sec:turbdisk}
We begin by studying how turbulence can drive clump formation using full turbulent disk simulations.  To create these cases, the entire disk is overlaid with a turbulent spectrum in a manner similar to \cite{price2010}.  We applied a Burgers' turbulent spectrum ($k^{-4}$) with a peak scale of $2$ kpc to the initially quiet disks from Section \ref{sec:methods}.  This gives an rms turbulent velocity of 7, 20 and 50 km/s in the cold, warm and hot disks, respectively. There is no stellar feedback to sustain the turbulence so it decays over a few crossing times.  Results at time 40 Myr are shown in the left column of Figure~\ref{fig:turb}, this is slightly more than twice the shear time ($\kappa^{-1}$). These disks are clearly visually clumpy, but we require a more quantitative measurement.

If we look closely at the galaxies in Figure \ref{fig:turb} there is also some  faint spiral structure present, particularly in the warm and hot disks. \cite{inoue2018a} and \cite{inoue2018b} have put forward a theory for massive clump and cluster formation from spiral arms. However, in our simulations it is not clear if the spiral arms drive clump formation or vice versa. For this reason, in the following sections, we do not draw explicit comparison between the models of \cite{inoue2018a}.

\subsection{Measuring Clump Masses} 
For cases that are able to produce a clump structure, we must be able to assign a mass to these objects.  We use two different approaches to assign clump masses.  To study clump masses in manner similar to radio frequency observations \citep[e.g.][]{colombo2014}, we use the package astrodendro \citep{rosolowsky2008} to identify massive, dense structures in our simulations.  As a second approach we use SKID (Spline Kernel Interpolative Denmax)\footnote{https://github.com/N-BodyShop/skid} to identify bound objects in our simulations.  We have chosen to identify "clumps", which could be the progenitors to star clusters, as bound objects. SKID is able to identify groups of bound gas particles and so is perfect for this purpose.

\subsubsection{Finding clump masses with Astrodendro}
  The package astrodendro uses dendrogram trees to identify related structures.  Dendrograms are particularly useful for identifying objects embedded in larger hierarchical structures, in a way that requires limited parameter choices \citep{rosolowsky2008, colombo2015}.

In order to pass data through the astrodendro package we convert all of our data into synthetic FITS files.  For this conversion we use the method of \cite{ward2012}, wherein particles are mapped into pixels and then smeared using the the SPH smoothing kernel.  To make the identification more straightforward we zoom in on a region surrounding the clump, with a width of 1 kpc.  For this analysis we use pixel sizes of 10 pc.  At this time we do not make any attempt to map the gas to CO emission, and instead use the surface density.

The only parameter we set in astrodendro is a minimum value required for a pixel to be considered part of the tree.  We have experimented with other parameters (mindelta, minnpix), but find that at our resolution they make little difference.  We experiment with two thresholds for the tree.  The first is a surface density of $100$ M$_{\odot}$/pc$^2$, above which we can estimate that star formation would proceed.  This is close to the extinction threshold of \cite{lada2010}.  The second is a surface density of 10 M$_{\odot}$/pc$^2$, above which gas transitions to being mostly molecular \citep{bigiel2008}.  We find that this threshold makes little difference to the masses of the objects found.  From this point on, we use the higher threshold of 100 M$_{\odot}$/pc$^2$ to build the trees.  

\subsubsection{Finding clump masses with SKID}
\begin{figure*}
	\includegraphics[width=0.8\textwidth]{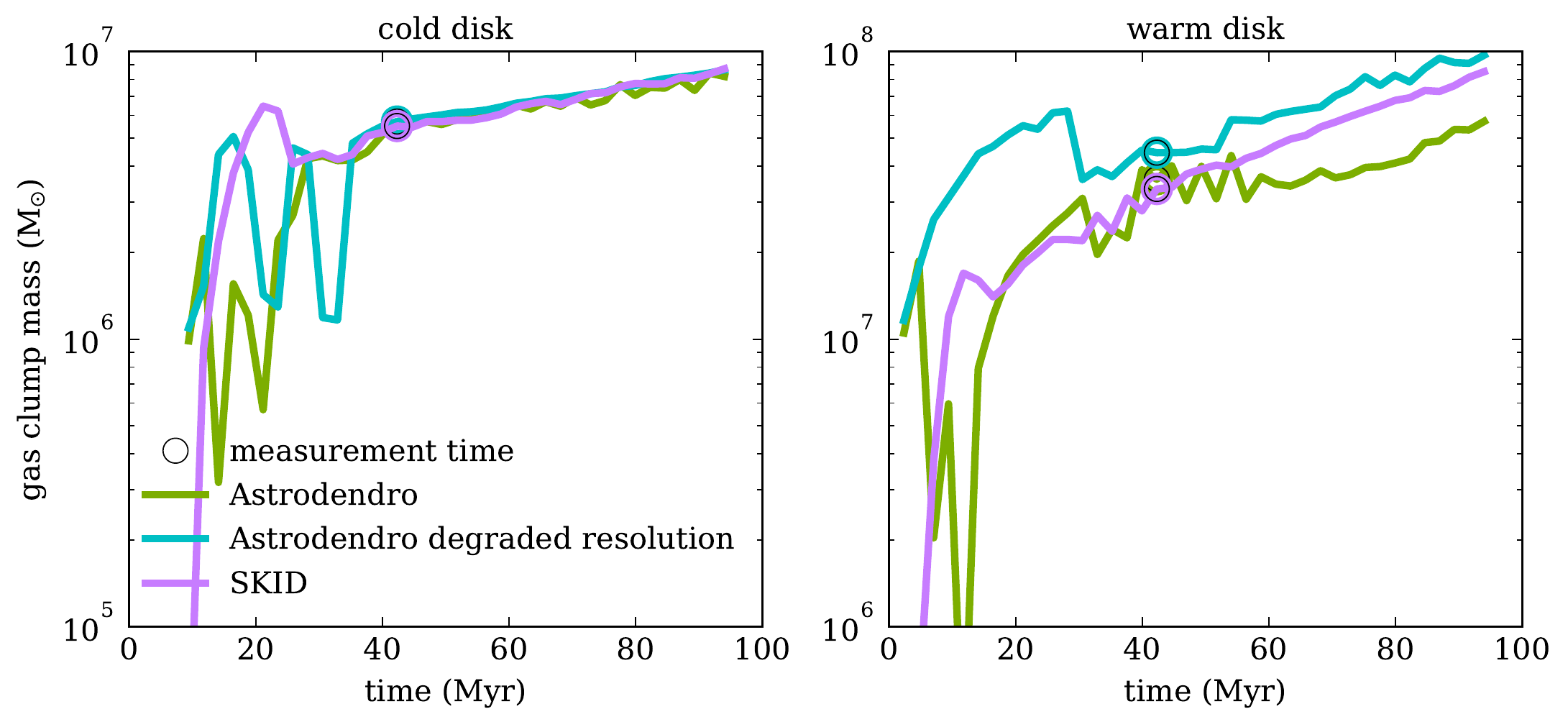}
	\caption{A comparison of the clump masses found by astrodendro versus SKID in the cold (left) and warm (right) disk.  The green line shows the clump mass found by astrodendro as a function of time.  The teal line shows the mass found by astrodendro if the initial resolution of the map is degraded to 100 pc.  The purple line shows the bound mass identified by SKID as a function of time.  The circles denote the time at which we record the mass measurement.  }
	\label{fig:astrodendro}
\end{figure*}
SKID is an Nbody group finder.  It identifies structures using a friends-of-friends algorithm.  Each structure then has unbound particles removed.  Thus the final group contains only the bound mass formed.  To find these masses with SKID the only essential parameter is a linking length.  We choose a group linking length of 20 pc.  Additionally, we consider only particles above 100 cm$^{-3}$, as these particles are dense enough to be part of the molecular medium. When recording the mass using SKID we take the mass of the central bound object. For the remainder of this section we report results using only SKID.  We defer a detailed comparison of SKID and AstroDendro to a later section.

\subsection{Clump masses in the turbulent disk}

Figure \ref{fig:turb} shows results for the cold, warm and hot disk. The top row of Figure \ref{fig:turb} shows the surface density maps while the bottom row shows the mass histograms for bound objects in the disk at the chosen measurement time. The Toomre mass is plotted in each histogram as the dashed line.  The maximum masses produced are $5.3\times10^6$ M$_{\odot}$,  $1.1\times10^8$ M$_{\odot}$ and $2.5\times10^8$ M$_{\odot}$ for the cold, warm and hot disk, respectively. At first pass, our results do not agree with our estimate of the Toomre mass.

When comparing simulations to analytic theory, assumptions and interpretations must be made. These interpretations manifest themselves in the transition from a theoretical length scale to a measurable mass. The Toomre length is the largest linear mode that can collapse for a given disk rotation profile. The actual largest scale of collapse is expected to be smaller; commonly used multiples include 1/2, 1/4 or even 1/16.

We may instead choose to compare to the fastest growing mode, or the critical wavelength, $\lambda_{\rm{Toomre}}/2$.  This leads to masses a factor of four smaller than the Toomre mass. This region is shaded in dark grey in Figure \ref{fig:turb}.  We may also assume that the entire unstable region does not contribute to the growth of the final bound object. For instance, \cite{dekel2009} assume that only the inner $\lambda_{\rm{Toomre}}/4$ collapses. This region is shaded in light grey in Figure \ref{fig:turb} and this lowers the mass estimate by an order of magnitude.

One can attempt to include turbulence as a stabilizing factor.  For example an effective Toomre parameter can be estimated by replacing the sound speed, $c_s$, with a combined term, $\sqrt{c_s^2 + \sigma^2}$. If we consider this higher amount of turbulent support our disks have initially higher Q values, around 1.5-1.7. Further, the current formulation we use for the Toomre length assumes that rotation is the maximum limiter. Intermediate between stabilizing contributions from pressure and rotation lies the most unstable scale:
\begin{equation}
\lambda_{\text{MU}} = \frac{2\sigma^2}{G\Sigma}.
\end{equation}
We find that with the amount of turbulence present in the disks $\lambda_{\text{MU}}$ is larger than our calculated Toomre length. For the cold, warm and hot disk these are 1.6 kpc, 4.6 kpc and 9.9 kpc, respectively, approaching the size of the disk itself.

However, it is not clear that turbulence should be treated in the same sense as thermal pressure in this context.  Thermal pressure contributes uniformly to the stability of the disk. Turbulence, however, is intermittent (i.e. may be absent in a specific region) and depends on the scale of measurement. Further, the wavelengths calculated above are far too large to act as limiters or predictors of structure formation. As we have seen in the simulations in this section, turbulence quickly leads to non-linear fragmentation.

The results depend on parameters of the turbulence, which are difficult to pin down. Additionally, the largest clumps are rare and are difficult to quantify using turbulent simulations due to small number statistics. In the following sections we will no longer consider our turbulent simulations.

\section{Seeding Clump Formation}\label{sec:seed}

Turbulent disks simulations have set an expectation for the range of clump masses possible in this work and others.  However, they are not an ideal comparison point, rather they have illustrated the complexity of the problem.  In particular, it is not possible to link each clump with the size scale and strength of the perturbation that seeded it. If we wish to compare to theoretical expectations we require a more controlled setup.

\subsection{Seeding clump formation in quiet galaxies}\label{ssec:seed}
In this section, we present a new approach to studying clump formation, or more generally the formation of bound structures, in simulations. In nature, turbulence is generated on large scales and then cascades down to feed smaller scales.  It is these smaller scales which are of key interest in star formation.  However, simulations struggle to capture the full turbulent cascade; it is difficult to resolve and maintain the full turbulent spectrum at the resolutions available for most galaxy-scale simulations \citep{kritsuk2007}.

In Section \ref{sec:turbdisk}, we avoided this problem by laying down a spectrum of turbulent velocities at the beginning of the simulation. 
The shape of our perturbation is modelled by the function

\begin{equation}
\Delta \vec{\text{v}} = -\text{v}_0\ \left(\frac{r}{l/4}\right) \ e^{-0.5(r^2/(l/4)^2)}\ e^{0.5} \cdot \hat{r},
\label{eqn:pert}
\end{equation}
where $r$ is the distance from the centre of the perturbed region, $v_0$ is the chosen perturbation velocity, and $l$ is the chosen perturbation wavelength. We choose this form such that $v_{max}=v_0$ at $r=l/4$, similar to a sine wave with wavelength $l$ but quickly returning to zero.  For this function, the divergence is almost uniform (and negative) for $r < l/4$ and then smoothly returns to zero.  Thus the characteristic collapse time is $l/(4 v_0)$. We would expect that in nature such a convergent flow may be caused by neighbouring superbubble events, as one example.

Effectively, we are introducing a radially compressive mode in the region of interest, such as could be generated by local feedback events. By then choosing different combinations of disk mass, perturbation wavelength and perturbation velocity, we can build a large parameter space of cases. In this way we can explore the conditions that led to bound structure formation in disks of different masses.

\subsection{Comparison of SKID and astrodendro}
The clump masses as found by SKID and astrodendro agree when compared, after 30 Myr (the differences in the warm disk are no more than a factor of two). To illustrate this agreement, we choose two of our seeded clumps and follow their masses as a function of time with both SKID and astrodendro. The results of this analysis are shown in Figure \ref{fig:astrodendro}. The left panel shows the results for a cold disk with a wavelength of 1 kpc  and perturbation velocity of 10 km/s.  The right panel shows the results for a warm disk with a wavelength of 1 kpc and a perturbation velocity of 30 km/s. The different coloured lines show different different clump finding methods; the purple line shows the results for SKID while the green line shows the results for astrodendro in each panel.  Additionally, the teal line shows the results for astrodendro with the resolution degraded from 10 pc to 100 pc.

There are significant differences in the identified masses, but these differences occur while the clumps are still forming. The important region to consider is the span of time after which an identifiable bound object is formed. As mentioned previously, we take our mass measurements after $\sim30$ Myr, this time is denoted as an open circle in Figure \ref{fig:astrodendro}. In the cold disk, after this time, there is less than 10\% variation between the different methods. As noted above, there are slightly larger differences when looking at the warmer disk, where the background surface density is intrinsically higher.

\subsection{Clump Evolution}
\begin{figure}
	\begin{center}
	\includegraphics[width=0.45\textwidth]{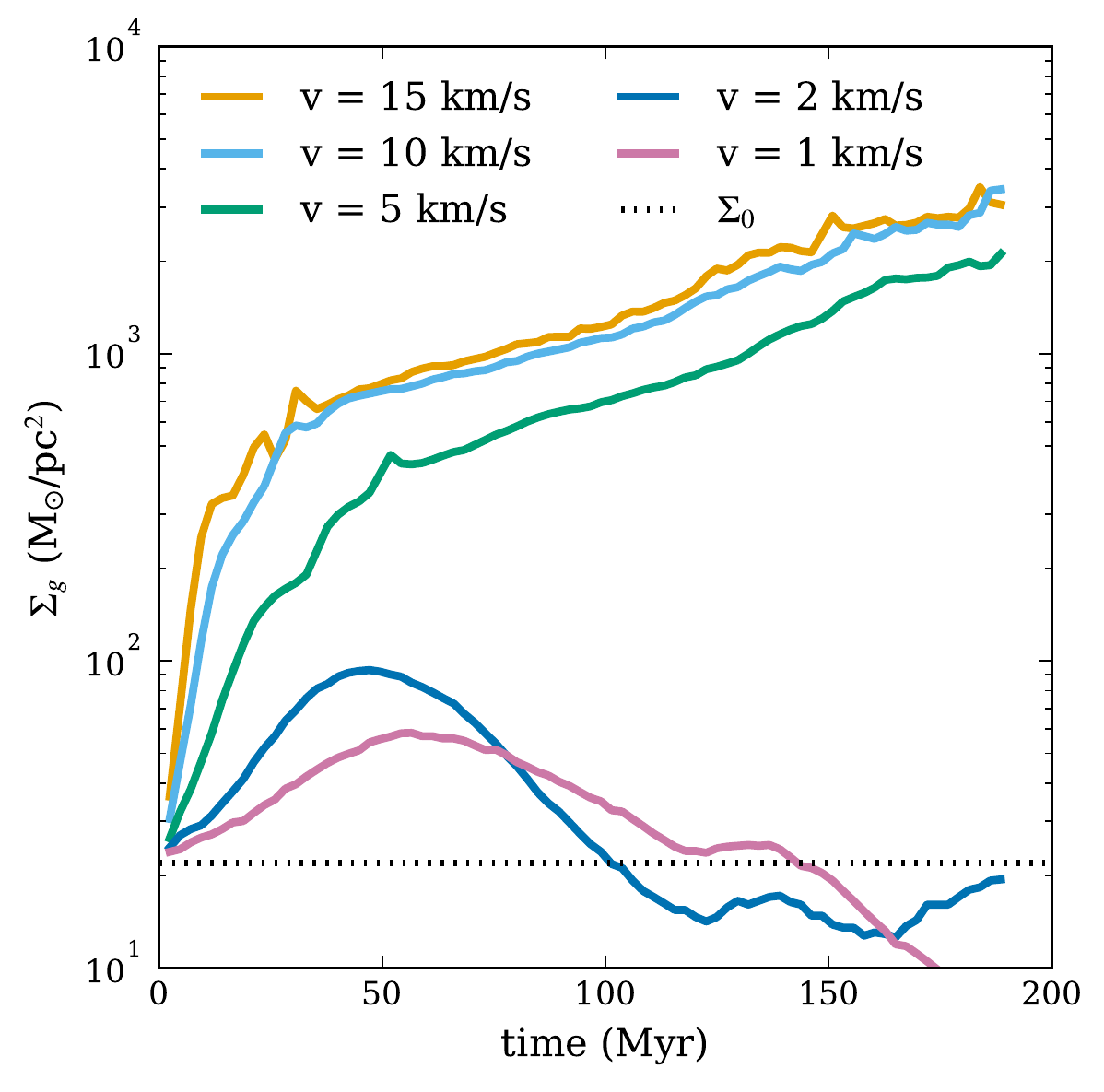}
	\end{center}
	\caption{The evolution of the surface density for perturbations in the cold disk.  The lines plot the surface density in a 50 pc aperture centred on the initially perturbed region.  The cases with the smallest velocities, 1 and 2 km/s, show examples of perturbations that failed and could not produce a clump.  The larger velocities, 5, 10 and 15 km/s, show examples of perturbations that produced a central clump.}
	\label{fig:sigmaholdl}
\end{figure}

As mentioned previously, for the purposes of analysis we define a "clump" as a bound gas structure. We further require that this structure formed as a result of one of our seeded perturbations. To identify cases that are able to grow structure, we track the surface density of the perturbed region. For each case, we track the surface density in a 50 pc apertures centred on the initially perturbed particles. A sample plot for the evolution of the surface density with time can be seen in Figure \ref{fig:sigmaholdl} for five cases. The cases in Figure \ref{fig:sigmaholdl} span a range of velocities, from 1 to 15 km/s, but the wavelength of the initial perturbation is held constant at 1 kpc.  The evolutionary tracks separate themselves into two distinct sets, those that increase in surface density and those that do not; we will discuss this in depth in the following sections. 

\begin{figure}
	\begin{center}
	\includegraphics[width=0.5\textwidth]{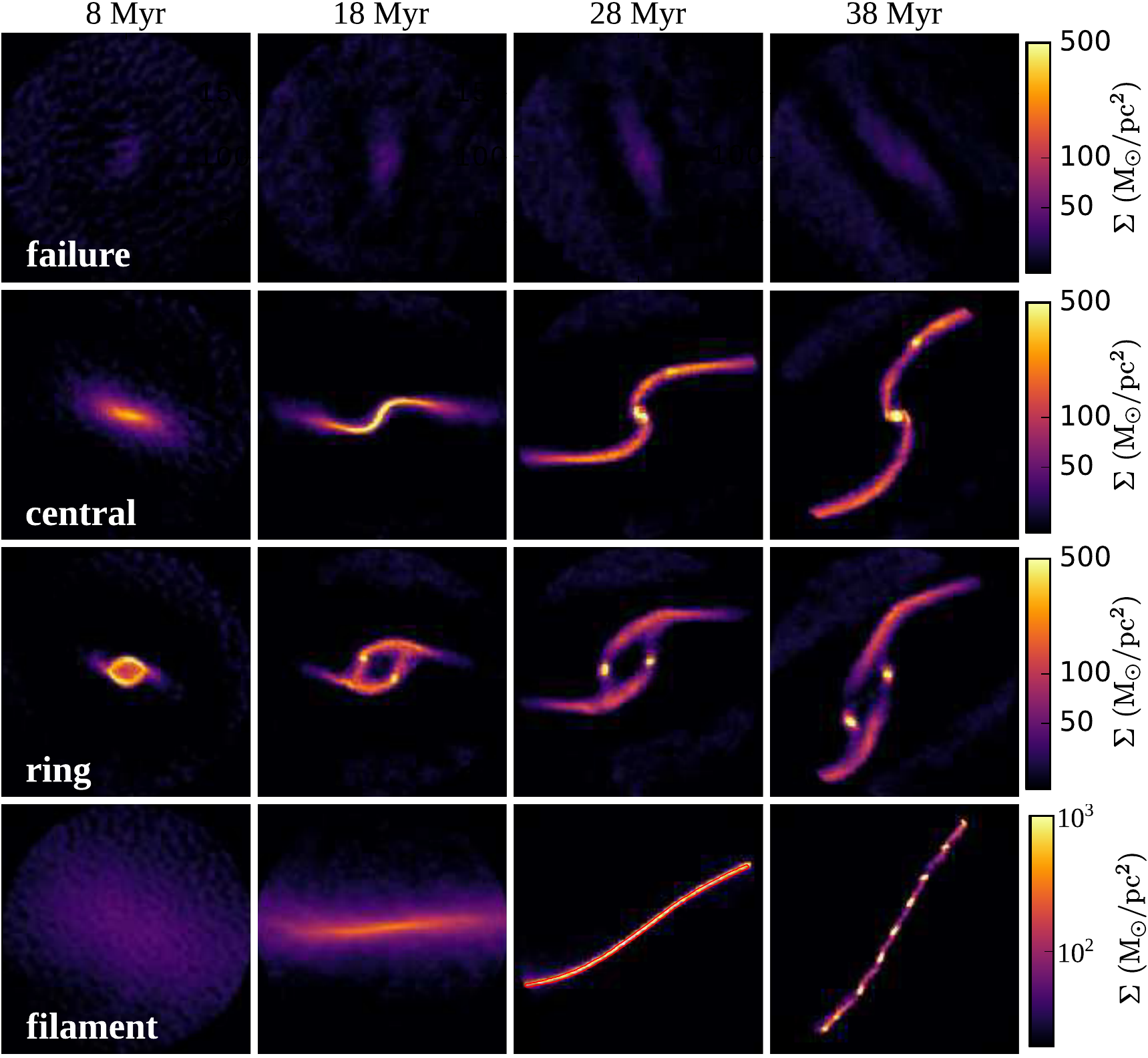}
	\end{center}
	\caption{Evolution of the surface density for different clump formation scenarios over time.  \textit{Top row}: a failed clump, as discussed in Section \ref{ssec:fail}. \textit{Second row}: a central clump, as discussed in Section \ref{ssec:central}.  \textit{Third row}: two ring clumps, as discussed in Section \ref{ssec:ring}.  \textit{Bottom row}: a filament, as discussed in Section \ref{ssec:filament}. The image axes denote the x and y dimensions in the plane of the disk. The direction to the centre of the disk varies from frame to frame. The snapshots are 1 kpc across.}
	\label{fig:cases}
\end{figure}

\begin{figure*}
	\includegraphics[width=\textwidth]{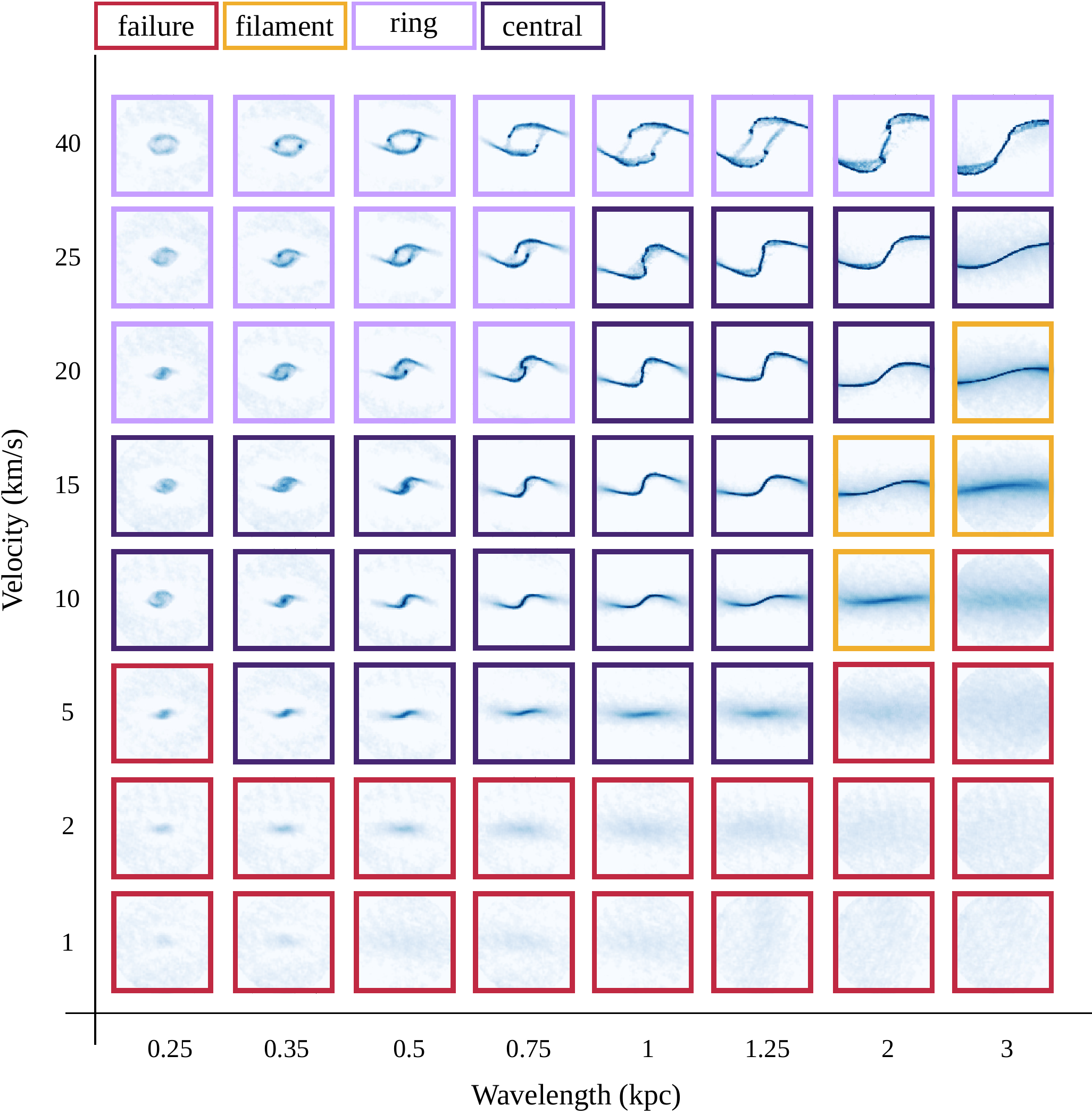}
	\caption{Sample surface density images of different perturbations in the cold disk.  There are four different types of cases when considering the evolution of the perturbation.  Weak perturbations, with low velocity are clear failures; these images are outlined in red.  Successful cases are those that form a dense object at their centre, where the original perturbation was centred.  These cases are outlined in dark purple and occupy a distinct region in the $l$-v space.  Strong perturbations, with high velocity, often lead to the formation of ring clumps.  These cases are outlined in light purple.  Lastly, are cases we label as filamentary.  These are cases that exceed the maximum mass per unit length of a filament for fragmentation.  These cases are outlined in yellow.  These snapshots are taken at the shear time for the disk, $\sim15.5$ Myr, and are 1 kpc across.}
	\label{fig:composite}
\end{figure*}

To help visualize what our clump evolution actually looks like, we have plotted sets of surface density maps in Figures \ref{fig:cases} and \ref{fig:composite}. For Figure \ref{fig:composite}, the snapshots are taken at the shear time,
\begin{equation}
t_{\text{shear}} = \frac{1}{\kappa},
\end{equation}
where $\kappa$ is the epicycle frequency for the disk.  In this case the shear time is 15.5 Myr. This is the time to shear out a distance equal to your size.

We choose the shear time as the time to take representative snapshots as the perturbations have their fate determined by this point. On this timescale, a sheared structure can be shifted by order its own size; it is the period of a full epicycle. In this disk there are two main ways for a perturbation to fail.  The first is thermal support due to the background sound speed in the disk. The second results from rotational support: this rotational support is associated with the initial shear and any later stretching due to ongoing shear. Thus, after one epicycle the perturbation should have started to collapse, or otherwise have begun to be sheared apart.  

For the remainder of this section we discuss only the cold disk and use it as a case study to illustrate key concepts.  All of the behaviour and cases discussed occur in all three of our galaxy disks, just in different regions of the $l-v$ parameter space.

\subsection{Failed Perturbations}\label{ssec:fail}
Referring back to Figure \ref{fig:sigmaholdl}, two of the cases, with velocities of 1 and 2 km/s, are examples of perturbations that were not able to grow. Initially, they experience an increase in surface density, however, by $\sim$ 50 Myr, both have turned over and decrease until they oscillate around the background surface density of the disk; they are stabilized by rotation and pressure. This behaviour can be seen in the top row of Figure \ref{fig:cases}, where a sample surface density map is shown, in time spacings of 10 Myr. After 38 Myr of evolution, right before we take a mass measurement, there is no dense structure present. Cases like these are deemed "failures" and, looking at Figure \ref{fig:composite}, they occur mainly in the low velocity range. We can further define these cases as those that experience a turnover in their surface density.

The bottom two rows in Figure \ref{fig:composite} show examples of failed cases, which are outlined in red.  In a direct analogy with the Toomre criterion for linear perturbations, clump collapse can be opposed by both rotation and pressure.   The relevant timescales are the shear time and the sound-crossing time.  Typically, depending on the scale, one of these dominates within a small region.  Within a region roughly half the Toomre length, they are comparably important.  This is quite similar to expectations for the linear, axisymmetric case (i.e. the Toomre criterion) and the non-axisymmetric case \citep{goldreich1965b}.  In both those scenarios, and in this work, it is always the case that pressure and rotation work together to dissipate structure.  Moving away from axisymmetry slightly boosts the role of rotation (see eqn \ref{eqn:glb}).

The failed cases in Figure \ref{fig:composite} trace out a rough parabolic shape. This parabola is reminiscent of the growth rate as a function of $k$ for the Toomre instability. In the pure axisymmetric case, rings cannot shear and so the shear time is not important.  In that case, any growth rate larger than zero leads to the growth of structure.  However, in our non-axisymmetric case, rings can shear.  This means that the growth rate must compete with the shear time: if the growth rate is not sufficient, perturbations will be sheared apart before structure forms.  In our study, this is the case for long wavelengths. For shorter wavelengths the competition comes down to the pressure, or the sound crossing time. At those wavelengths, the sound crossing time must be less than the collapse time. In the coldest disk, this effectively requires that the perturbation velocity be larger than 5 km/s.

\subsection{Central Clumps}\label{ssec:central}

Cases where clumps have formed in the centre of the initially perturbed region are the easiest to interpret.  They can easily be identified because the evolution of their surface density shows a rapid phase of increase, followed by a phase of constant, steady growth.  In Figure \ref{fig:sigmaholdl} the cases with velocities of 5, 10 and 15 km/s are perfect examples of cases that form central clumps.

We note that at later times, when the surface density has climbed too high, these clumps are numerically poorly resolved.  We know that they are bound and expected to collapse further. This has been shown analytically by \citep{goldreich1965b} for non-linear collapsed perturbations with an isothermal equation of state.  However, the precise evolution of the surface density with time would require adaptive resolution beyond that employed here.  For that reason, we do not place any special meaning on the steady exponential growth phase seen for collapsed objects past $\sim 50$ Myr (as is seen in both surface density and mass, as shown in Figures \ref{fig:astrodendro} and \ref{fig:sigmaholdl}).

We are also able to identify trends between the initial perturbation parameters and the resulting surface density of the central region.  We see that as the initial velocity is increased, within a shear time, it is possible to get higher and higher surface densities.  This effect does begin to saturate if the velocity gets too high, above $\sim$ 15 - 20 km/s (see next section).

The second row of Figure \ref{fig:cases} shows an example of the evolution of a central clump.  There is a stark difference when this case is compared to the surface density of the failed case.  For a central clump, already by 8 Myr the surface density in the middle of the perturbed region has surpassed 100 M$_{\odot}$/pc$^2$ and by 28 Myr there is clearly a dense bound structure present.  In Figure \ref{fig:composite} these central collapse cases are deemed "centrals" and outlined in dark purple.  They occupy a specific region of the $l-v$ space.

For this scenario of clump formation, in the cold disk, masses between $8.47\times10^5$ M$_{\odot}$ and $1.6\times10^7$ M$_{\odot}$ are possible.  All the results for our measured clump masses are plotted in Figure \ref{fig:mass}, where the top panel shows the cold disk we discuss in this section.  Clumps that form by fragmenting in the centre of the perturbation, as discussed here, are plotted as filled symbols.  In Figure \ref{fig:mass} we can see general trends with both velocity and wavelength.  Generally, as the velocity is increased, the masses of the central clumps also increase.  Similarly, as the wavelength of the perturbations is increased the masses of the clumps formed tend to increase as well.  

\subsection{Ring Clumps}\label{ssec:ring}

A clump forming in the centre of a perturbation in the location where all of the velocities were directed is the most straightforward situation.  However, there are more complicated scenarios that lead to the formation of bound structures.  One such scenario involves the fission of a ring-like structure.  The perturbations are initially directed radially inward.  The higher the initial velocity, the larger the radius from which material can reach the centre quickly, within the shear time, for example.   The angular momentum of the initial disk material with respect to the centre of the perturbation increases as $r^2$, where $r$ is measured from the perturbation centre.  Therefore, for large velocities the  angular momentum per unit mass of the initially collapsed clump becomes quite high.  This high angular momentum can cause the centre region to have too much rotational support and thus re-expand as a ring.

This ring expands and, as it does, is caught in the large-scale shear flow.  This means that portions of the ring can be dragged spinwise or counter-spinwise until they are ultimately stretched out to join tidal-like features.  The rings thus have strong m=2 type modes, which results in two over-densities rather than one. 

The third row of Figure \ref{fig:cases} shows an example of the evolution of two ring clumps.  Here again by 8 Myr the surface density has exceeded 100 M$_{\odot}$/pc$^2$, this time in the ring structure.  By 20 Myr, the two ring clumps have begun to fragment.  By 38 Myr, the ring has begun to be sheared apart, and the two ring clumps are clearly visible. Fragmentation in high spin cases leading to multiple objects has been found to occur in protostellar disks \citep{kratter2006,kratter2008}.

Since this scenario requires a large angular momentum per unit mass in the centre of the perturbed regions, these cases are seen at high velocities.  In Figure \ref{fig:composite} the ring cases are outlined in light purple and can be found for cases with v $\gtrsim 20$ km/s.  For the final masses in these cases we take the total bound mass of the two ring clumps.

The clumps formed from rings tend to have similar masses to central clumps when the two bound structures are counted together.  In this case, total masses of between $9\times10^5$ M$_{\odot}$ and $1.8\times10^7$ M$_{\odot}$ are typical in the cold disk.  When considered individually though, the clumps are less massive than those formed through the central fragmentation scenario.  In Figure \ref{fig:mass} the open symbols show the masses of ring clumps.  Generally, these clumps follow the same trends as the central clumps.  As the velocity is increased the clump mass increases, and as the wavelength is increased the clump mass increases.  We can also see that, in general, the ring clumps have total masses that are higher than the central clumps. These high spin objects are consistent with the idea that observed large objects could be the result of beam crowding. In that case, we would actually be seeing the mass of multiple clumps in close proximity that formed out of a similar structure.

\subsection{Filamentary Fragmentation}\label{ssec:filament}

As noted earlier, when considering non-axisymmetric perturbations and long enough timescales, shear dominates the evolution of everything except the innermost bound clump.   The shear will eventually draw the unbound material out into ever elongating tidal arm features.  In an unrealistically quiet disk (such as in the controlled cases here) and over very long timescales, this material is wrapped to form a ring-like structure.  In practice, the galaxy would have other perturbations acting on smaller timescales that break the the material up and use it to form other structures.   However, there is an intermediate stage where the features are drawn out into filaments. 

Perturbations with long wavelengths can exhibit a different mode of fragmentation.  As time passes the long structures begin to behave like star-forming filaments.  Such a filament will begin to fragment once it's mass exceed the critical line mass,
\begin{equation}
M_{\text{line}} = \frac{2c_s^2}{G}
\end{equation}
where $c_s$ is the sound speed and $G$ is the gravitational constant \citep{inutsuka1997}.  For our coldest disk the line mass works out to 11,624 M$_{\odot}$/pc.  The average width of our long wavelength filaments is on the order of 40 pc.  This means that our filaments only have to exceed surface densities of $\sim$290 M$_{\odot}$/pc$^2$ to be above the critical line mass and begin to fragment. 

As time goes on and shear continues to operate, the filaments become more and more drawn out, decreasing the mass per unit length.  This suggests there is a set time limit for the fragmentation in these objects to begin.  If the original density of the filament is not high enough they will just be sheared apart.

We see this exact behaviour in many of our long wavelength perturbations.  The bottom row of Figure \ref{fig:cases} shows an example of a perturbation that develops filamentary behaviour and then fragments.  By 18-28 Myr the density in the filament has begun to exceed the critical surface density calculated above.  Indeed, by 38 Myr the filament has completely fragmented.

More examples of these filamentary cases can be seen on the right side of Figure \ref{fig:composite}, outlined in yellow. We label any structure produced by this phenomenon as secondary fragmentation.  While it does lead to bound structures, there is different physics responsible for their production on top of our initial perturbation.  In general these fragments are less massive than the central or ring clumps for comparable wavelengths and velocities.  For example, at a wavelength of 2 kpc and a perturbation velocity of 10 km/s, the filamentary fragments are on average $2.4\times10^6$ M$_{\odot}$.  As noted above, only very quiet disks would get the opportunity to use this mode of fragmentation so we have elected not to use it when comparing to observed disk properties.  These masses are not included in Figure \ref{fig:mass} or in any of the following discussion of mass.

\begin{figure}
\begin{center}
	\includegraphics[width=0.45\textwidth]{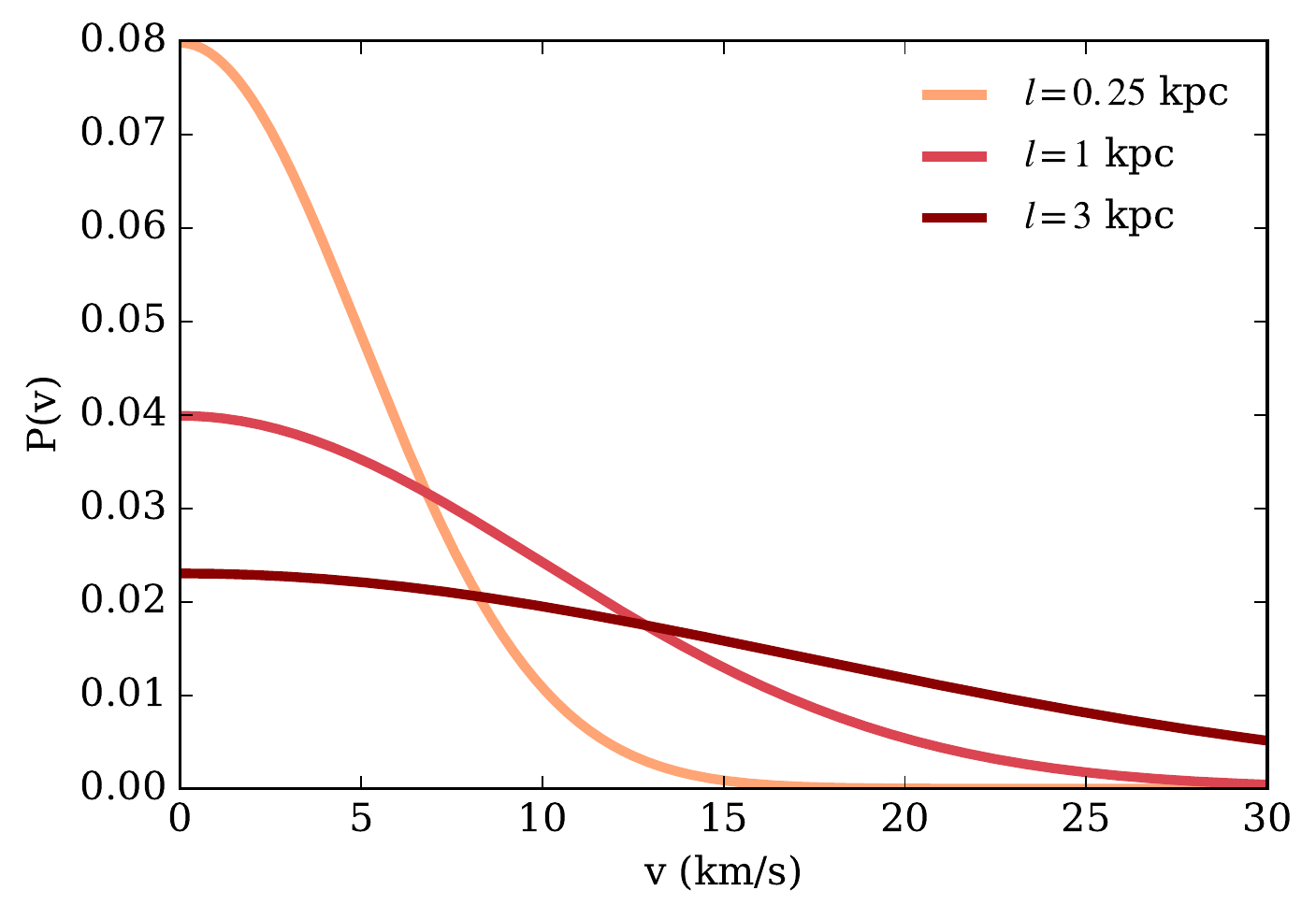}
	\end{center}
	\caption{Sample probability distributions for three different wavelengths in the cold disk.  These are generated assuming a spectrum of the form shown in equation \ref{eqn:turb}. These are examples of the spectrums used to generate the most likely mass for each wavelength.  As we can see from the plot, for the cold disk. velocities above 15 km/s are very unlikely.}
	\label{fig:gaussian}
\end{figure}

\section{What is the most likely clump mass?}\label{sec:mass}

In this section we use our measured clump masses to determine the most likely clump mass in a given galaxy.  We begin with the coldest disk in our sample.  In nature, bound structures, like clumps, are seeded by a cascade of turbulent energy.  An extensive amount of work has been done to characterize the nature of this turbulence in molecular clouds \citep{myers1983, solomon1987, kritsuk2013}.  There is a correlation between the velocity dispersion, or linewidth, and the size of a region \citep{larson1981}.  Larson's first relation states
\begin{equation}
\sigma = A \, l^{1/2},
\label{eqn:turb}
\end{equation}
where $\sigma$ is the velocity dispersion, $A$ is the normalization, and $l$ is the scale at which that velocity is generated.   For the cold disk, we assume that turbulence is generated by superbubbles, as it is in Milky Way-type galaxies.  This gives $\sigma_0$ = 10 km/s, which is generated on a scale on the order of the scale height of the galaxy, $l_0$ = 1 kpc.  

In order to assign a likelihood to each of our clump masses, we assign a probability to the velocity that seeded it.  To assign this probability, we next assume that all of the velocities at a given wavelength are drawn from Gaussian distributions.   Samples of the Gaussian distributions are shown in Figure \ref{fig:gaussian} for three different wavelengths; 0.25, 1 and 3 kpc. Generating turbulent velocities by drawing from a Gaussian distribution is common practice \citep[e.g.][]{price2010}.

\begin{figure*}
\begin{center}
	\begin{tabular}{c}
		\includegraphics[width=0.52\textwidth]{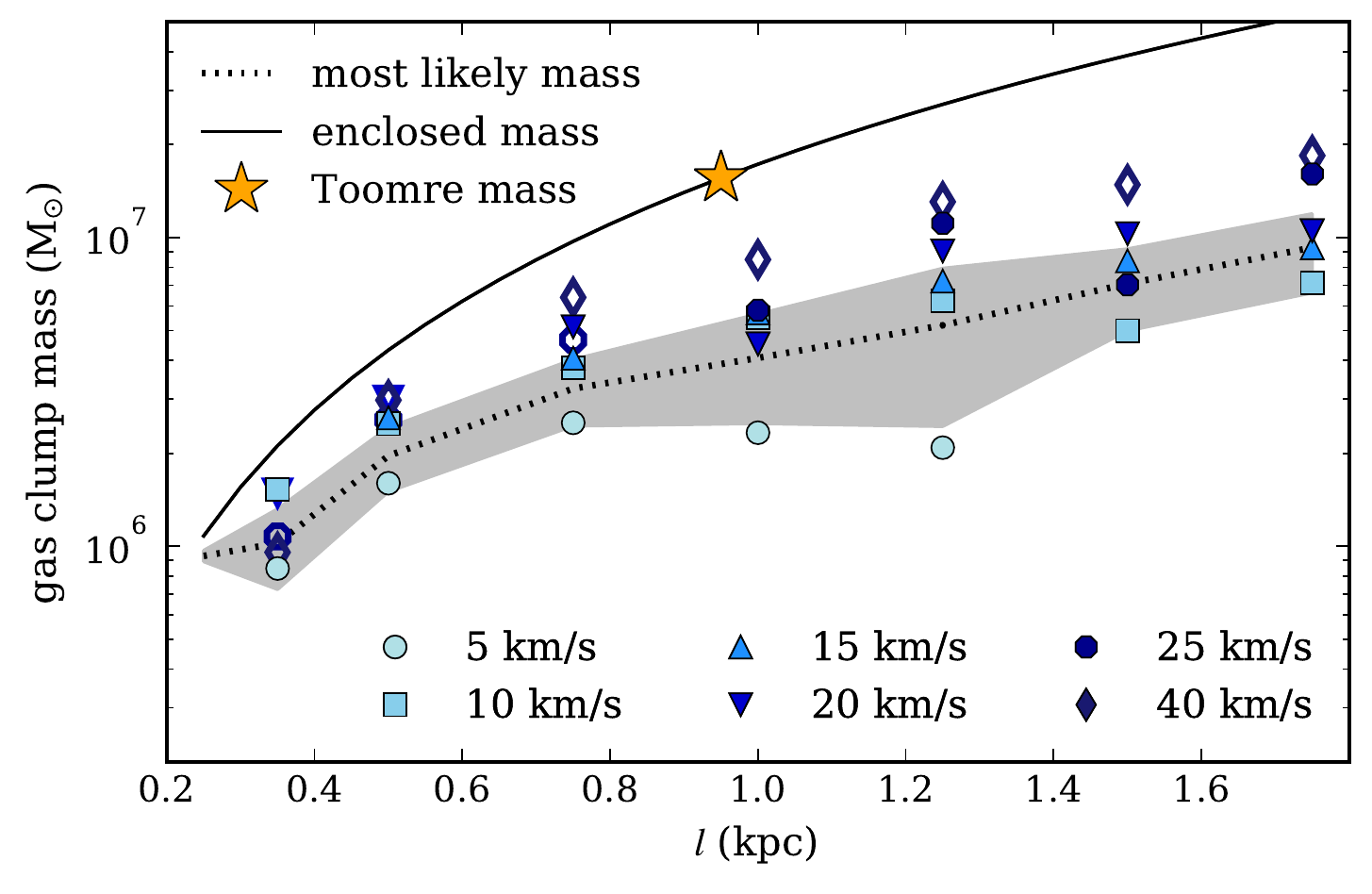} \\
		\includegraphics[width=0.52\textwidth]{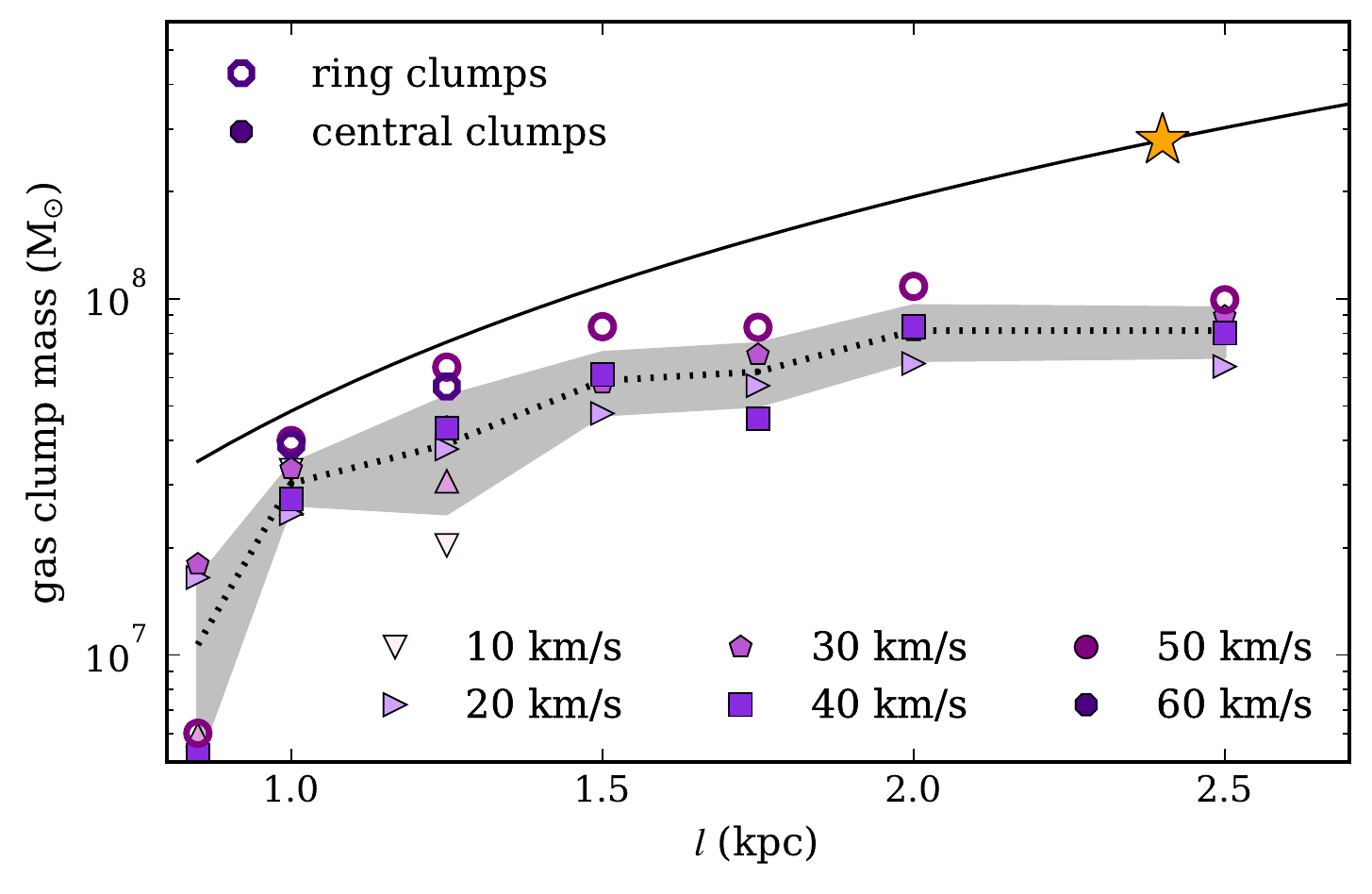} \\
		\includegraphics[width=0.52\textwidth]{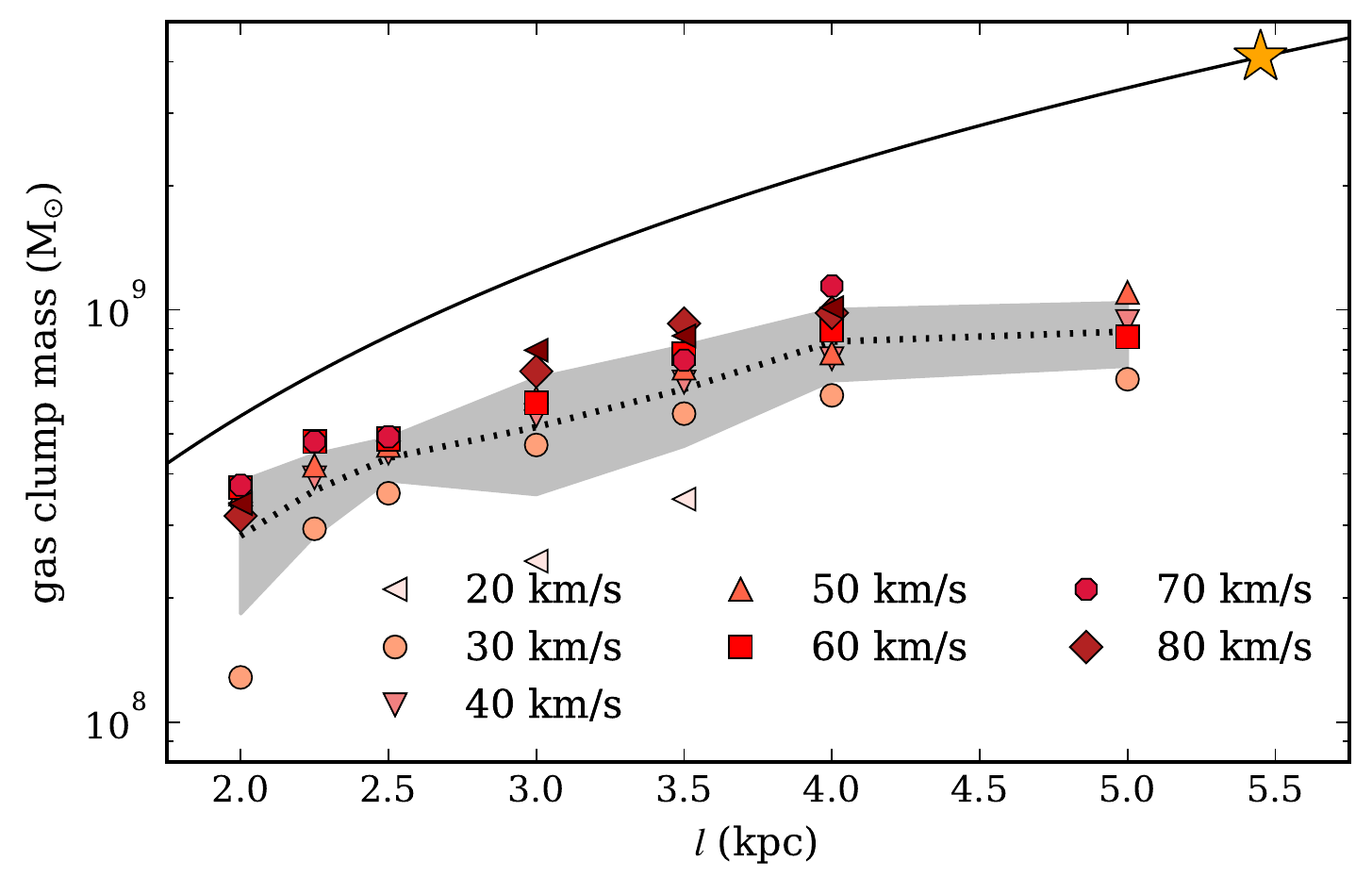} \\
	\end{tabular}
	\end{center}
	\caption{The most likely clump mass.  \textit{Top}: the masses in our cold disk.  \textit{Middle}:the masses in our warm disk.  \textit{Bottom}: the masses in our hot disk.  In each plot the filled symbols denote central clumps, the open symbols denote ring clumps and the stars show the Toomre mass at each wavelength.  The grey dashed line shows the mean expected mass and the shaded region shows the 1$\sigma$ deviation (see section \ref{sec:mass}).  Our measured clump masses wind up being smaller than the predicted Toomre mass.  The shaded $1\sigma$ region also shows us that, while it is possible to get closer to the Toomre mass, it is not a likely case in nature.}
	\label{fig:mass}
\end{figure*}

In general, coherent velocities on large scales above 10 km/s begin to get more and more unlikely.  By convolving these probabilities with the clump mass distributions that result from each velocity we can determine the most likely mass at each wavelength.  The results of this analysis are shown in the top panel of Figure \ref{fig:mass}. In the middle panel of Figure \ref{fig:mass} we plot the mass distribution for the warm disk, and the hot disk in the bottom panel.   For these cases we assume a velocity dispersion of $\sigma_0=50$ km/s \citep{wisnioski2014}.  For the generation scale, we assume that this level of turbulence would be generated on a disk scale length, on the order of 3 kpc.

It is important to note that this study is useful for providing upper bounds on clump masses formed via fragmentation, since we do not include any star formation or feedback and do not account for any late stage accretion.  That being said our approach offers a way to identify the most likely space for clump formation to operate in a given galaxy.  In Figure \ref{fig:mass} the dotted black line shows the most likely mass at each wavelength, while the grey shaded regions shows the 1$\sigma$ probable region.  We can see that in the cold disk, it is possible to get objects that begin to approach the predicted Toomre mass.  However, the velocities required to form these objects are very high; they are not probable velocities.  This in turn implies that while it is possible to make these larger bound structures approaching $10^7$ M$_{\odot}$, it is not a likely outcome.  This remains true as the disk mass, and thus the clump mass, is increased.

We can identify trends when considering the plots in Figure \ref{fig:mass}.  First, as the velocity of the perturbations is increased, the mass of the resulting clumps increases.  This is expected based on the the same trend observed for surface density in Figure \ref{fig:sigmaholdl}.  Second, there is a preferred mass scale for objects in a given disk.  For the coldest disk the preferred fragmentation mass lies around $3\times10^6$ M$_{\odot}$, and little variation is seen away from this.  This preferred mass lies around $4\times10^7$ M$_{\odot}$ and $5\times10^8$ M$_{\odot}$ for the warm and hot disks, respectively.  As we increase the disk mass, of course, the clumps that form are generally more massive.

Additionally, there is a preferred fragmentation length in each disk.  This length increases as the disk mass is increased.  It is not possible to form bound structures below 0.25 kpc, 0.85 kpc, and 2 kpc for the cold, warm and hot disk, respectively.   The physics behind these cutoffs was discussed in the previous section.

\subsection{The Toomre Mass}

As we did for the turbulent disk, we can compare the measured clump masses to the Toomre mass. As a reminder, for the coldest disk in our sample the critical wavelength is 940 pc, while the Toomre mass is $1.5\times 10^7$ M$_{\odot}$.  The Toomre mass is plotted as the orange star in Figure \ref{fig:mass}.  The solid black line shows what the enclosed mass would be at different wavelengths.  For the cold disk there is almost an order of mass discrepancy between the Toomre mass, and the expected mass (dotted line) at the critical wavelength.  As the wavelength increases, so does the discrepancy between our measured masses and the masses predicted by Toomre theory.
 
\begin{figure*}
\begin{center}
	\includegraphics[width=0.6\textwidth]{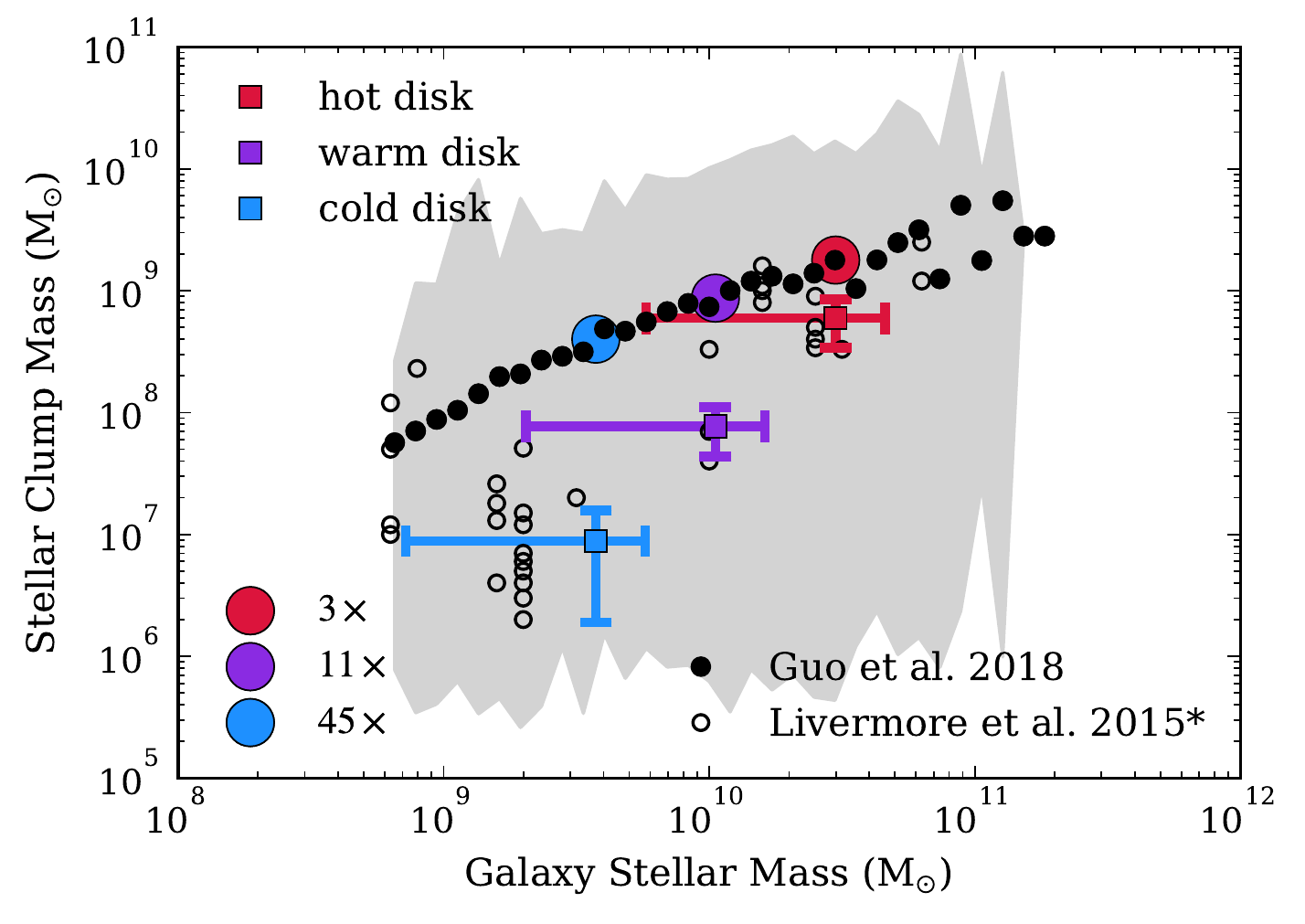}
	\end{center}
	\caption{The expected clump stellar clump mass as a function of the stellar mass content of the galaxy.  The average of our seeded disk clumps is plotted as each of the filled squares.  The vertical errorbars show the standard deviation of the data.  The horizontal errorbars show possible values of gas fractions for the conversion from gas to stellar galaxy mass, as well as account for the average sizes of observed galaxies.  Two observational samples are plotted for comparison. The black circles shows the average clump mass in 0.5 dex mass bins for the sample of \citet{CANDELS2} and the grey shaded region shows the range spanned by the data.  The open circles show our estimate of the clump mass for the lensed sample detailed in \citet{livermore2015}.  To convert from the original SFR measurements of \citet{livermore2015} we assume that a clump's lifetime is 100 Myr, and that the measured SFR is sustained for this entire time.  The large red, purple and blue filled circles show the level of beam crowding we would require to agree with the average CANDELS clump in a galaxy of comparable mass.}
	\label{fig:obs}
\end{figure*}
 
 In the warm disk, the critical wavelength is 2.4 kpc, which gives a Toomre mass of $2.9\times10^8$ M$_{\odot}$.  If we move to the hot disk those get even larger, with a Toomre mass of $4.1\times10^9$ M$_{\odot}$ resulting from a critical wavelength of 5.4 kpc.  These results are summarized in Table \ref{tab:summary}.  Again, we see that masses predicted from Toomre theory are consistently over-estimates when compared to our seeded clump masses.

We find that our masses are between a quarter and half of the expected Toomre mass. While the Toomre prediction is linear in construction, as objects grow more dense and fragment their behaviour is increasingly non-linear.  This is exacerbated by taking the results of a 1-d planar analysis, and translating it to a 2-d estimate of the mass. For these reasons, it is not entirely clear if discrepancies result from the non-linear nature of fragmentation or the application of that theory to mass predictions. Secondly, the system, and few systems in nature, are axisymmetric.  Finally, as the mass of the disk and its sound speed increase, the critical wavelength likewise increases.  In our most massive, hottest, disk the critical wavelength increases to 5.4 kpc.  A critical wavelength that large represents a significant portion of a galaxy disk.  In fact, at high redshift, that may be the entire radius of the galactic disk.  This has likewise been noted by \citet{reina2017}.  Through angular momentum considerations alone this scenario is not possible: the collapse of the whole disk is ruled out and thus $l \gtrsim r_d$ must fail.  For these reasons the Toomre mass should be treated as an absolute maximum mass, rather than a typical or even achievable upper mass limit.  

\section{Discussion \& Observational Implications}\label{sec:implications}

Up to this point we have assumed very fine resolution for our measurements.  This makes our mass measurements most comparable to high resolution work, like that of \cite{johnson2017}, with effective resolution of 40 pc.  To compare to a broad range of observational results we must try to account for the possibility of coarser resolution.

We choose 1 kpc, to be comparable to the beams in the CANDELS survey which are typically between 500 pc to 1 kpc.  This is in the range of resolution for multiple other high redshift surveys.  To compare to the stellar mass of clumps, we also estimate the location of newly formed stars in our disks.  We do not employ star formation and stellar feedback in these simulations.  However, we can tag likely locations for star formation by identifying all gas that surpasses 116 M$_{\odot}$/pc$^2$; this is the threshold identified by \cite{lada2010} for star formation. The simulations we use for comparison in this section are the seeded disks, discussed in section \ref{sec:seed}.

To estimate the impact of beam smearing we then change the way we assign masses to our seeded clumps.  Up until this point, the masses quoted are the masses of a single bound structure at the centre of the perturbation or two bound objects in the case of a ring structure.  In this section, we take all bound structures identified in the inner 1 kpc around the perturbation centre.  The results of this analysis are plotted as the squares in Figure \ref{fig:obs}.  These points show the average of the masses found in each disk.  The error bars show the standard deviation, found using the velocity averaging method discussed in section \ref{sec:mass}.

We have had to make assumptions about the stellar mass content of our disks.  We are using isolated galaxies in static potentials so our sample is not indicative of galaxies that have high stellar mass content.  Typical gas fractions for galaxies above redshift 0.5 can fall anywhere between 0.2 and 0.8 \citep{morokuma2015}.  Here we take the definition that the gas fraction is
\begin{equation}
f_{\text{g}} = \frac{M_{g}}{M_g + M_*},
\end{equation}
where $M_g$ and $M_*$ are the total gas and star mass of the galaxy, respectively. Again, since our galaxies cannot capture the dynamics of those with a high stellar fraction, we take the maximum possible gas fraction for our galaxies to be 0.5.

For comparison in Figure \ref{fig:obs} we plot the sample of clumps discussed in \cite{CANDELS2}. The clump stellar masses are quoted from Table 4 of \cite{CANDELS2}, which used the fiducial background subtraction algorithm.  The averages of the clump masses in galaxy mass bins of width 0.5 dex are plotted as the filled black circles.  The grey shaded region shows the range of the minimum and maximum clump mass in each bin.  We also plot the higher resolution lensed clump sample discussed in \cite{livermore2015} as the open circles.  We note that the sample of \cite{livermore2015} originally states only star formation rates for their clumps.  We have converted these to rough masses by assuming the clumps form stars at this steady rate for their estimated lifetime. 

The clump lifetime is still a debated topic.  Some theories suggest that clumps migrate into the centre of galaxy disks, which suggests their lifetimes must be longer than 150 Myr \citep[e.g.][]{shibuya2016}.  Other studies suggest a quick disruption due to strong feedback, giving a shorter lifetime of around 50 Myr \citep[e.g.][]{krumholz2010, oklopcic2017}.  We take a clump lifetime of 100 Myr to convert the sample.  This value means our masses estimated from the \citet{livermore2015} sample will be an underestimate for the mass if the clump lifetime is long, and an overestimate if the lifetime is short in nature.

We note also that our measurement of the disk stellar mass is influenced by our initial conditions. To avoid contamination from the edge or centre of the disks, we have seeded our clump formation events at $R = 5$ kpc in the disk. In nature, for $z>1$, typical effective radii for disks can be anywhere between 1 and 8 kpc, with a median around 2-3 kpc and strong redshift dependance \citep[e.g.][]{wuyts2011, vanderwel2014}. Our choice of seed radius is on the large end of these typical sizes. We can estimate the impact of this choice by imposing different truncation radii for our disks in determining the total gas mass. For example, if we take the disk mass at $R < 2$ kpc our mass estimate changes by less than an order of magnitude. This difference is already taken into account by the errorbars on the plot in Figure \ref{fig:obs}. Importantly, the only place where this makes a difference to our interpretation is on the high mass galaxy end, where the agreement between our results and observations were close to begin with. We stress that this only impacts the estimated stellar mass of our disks, and thus only the comparison made in Figure \ref{fig:obs}.

On the lower galaxy mass end, for our cold and warm disks, our sample agrees quite well with the findings of \cite{livermore2015}.  For our most massive galaxy, the hot disk, our clump masses agree well with both the \cite{CANDELS2} and \cite{livermore2015} samples.  As further comparison, the large blue, purple and red circles in Figure \ref{fig:obs} show how many of our clumps would have to be crowded in a beam to get our mass to the average \cite{CANDELS2} mass.  For the cold, least massive disk, we would need to have 45 of our clumps in the beam.  This number becomes more reasonable as we move to higher disk masses, 11 clumps in the warm disk and only 3 in the hot. These smaller factors agree well with the the resolution scaling factors discussed by \citet{cava2018}.

Our results suggest insights into clump formation, particularly in galaxies on the lower mass end (below $10^{10}$ M$_{\odot}$). We stress that the observationally measured mass of clumps is not necessarily the mass at which they formed.  Our results suggest that beam crowding likely plays a role in the measurement of large clump masses in these smaller galaxies. Alternatively, clump mergers may play a role in building larger objects. \citet{tamburello2015} also find that clump-clump mergers are required to build the largest objects. If clumps are long-lived, late-stage accretion can also build clump masses. On the higher disk mass end there are other ongoing effects that we do not discuss nor attempt to capture here. However, our results suggest that the largest observed clump masses likely result from the crowding or clustering of smaller objects into larger observational beams.

However, these results also come with caveats. First, we stress that our results should only be used as upper bounds, as we do not include the effects of additional fragmentation, variable star formation efficiency or stellar feedback on cluster mass. Second, we present here a very limited sample. We have simulated only three different types of galaxies, all of the same size and all with similar $Q$.  To firm up these results we would need a larger galaxy sample, spanning different sizes, surface densities and rotation curves. A future direction for this work would be explicit modelling of galaxies in the observational literature. We leave this to future papers.

\section{Summary \& Conclusions}

In this work, we have introduced a new method for studying the formation of clumps, or bound structures, in galactic disks. We seed clump formation events in initially stable isothermal disks, without star formation or feedback. We design these conditions to be purposefully simple and thus offer maximum control. Our clump mass spectrums are not impacted by feedback recipe choices, providing a complementary approach to other recent work that employs a variety of feedback assumptions. Our results indicate that the characteristic length and mass scale of clumps is not a fixed fraction of the Toomre estimate but depends on the properties of the galaxy in question. This complicates efforts to analytically estimate a characteristic mass.

By seeding turbulent clump formation events, we are able to study the exact conditions under which different clump masses form. In general, we find that our largest clump masses can be over an order of magnitude smaller than the Toomre mass. Our results suggest smaller initial masses for clumps than reported in some observational studies. This is consistent with the idea that those studies have large beams that encompass many bound objects.  We stress that when making this comparison, the clump observational mass may be different than the initial mass: just because an object is observed to be massive, does not mean it formed at that mass through gravitational fragmentation.

Our method provides a new way to approach the problem of studying clump formation in simulations; it offers a new way to compare to observations.  The method can be completely tailored to specific galaxies.  The only requirements is that the rotation curve and surface density distribution for the galaxy are known. Our method provides a promising new way to study the formation of these clumps in specific galaxies without the biases introduced by including different feedback methods.

\section*{Acknowledgements}

The authors thank Chelsea Sharon and Ralph Pudritz for helpful discussions.  SMB acknowledges support from the Vanier Canada Graduate Scholarship Program.  JW and HMPC acknowledge support from NSERC. This research made use of astrodendro, a Python package to compute dendrograms of Astronomical data\footnote{https://dendrograms.readthedocs.io/en/stable/}.  This research made use of Astropy, a community-developed core Python package for Astronomy (Astropy Collaboration, 2018)\footnote{http://www.astropy.org}.  Computations were performed on the GPC supercomputer at the SciNet HPC Consortium. SciNet is funded by: the Canada Foundation for Innovation under the auspices of Compute Canada; the Government of Ontario; Ontario Research Fund - Research Excellence; and the University of Toronto.


\bibliographystyle{mnras}
\bibliography{clumps} 





\bsp	
\label{lastpage}
\end{document}